\documentclass[aps,jmp,amsmath,amssymb,reprint]{revtex4-1}

\usepackage{graphicx}% Include figure files
\usepackage{dcolumn}% Align table columns on decimal point
\usepackage{bm}% bold math
\usepackage{graphicx}
\usepackage{subfigure}
\usepackage{physics}
\usepackage[english]{babel}
\usepackage[usenames, dvipsnames]{color}

\usepackage{hyperref}
\usepackage[normalem]{ulem}
\usepackage[utf8]{inputenc}
\usepackage{float}
\begin{document}

\bibliographystyle{apsrev.bst}
\renewcommand{\figurename}{Fig.}

% to remove the black line before reference
%\def\bibsection{\section*{\refname}} https://www.overleaf.com/project/632302f17bd9e44010bcf5b7

%\title{Calculation of electron transport parameters in the 2D semiconducting MXene $\bf{Ti_2CO_2}$ through an $\bf\emph{ab~initio}$ transport model for high-frequency applications}

%\title{Investigation of the high-frequency transport behaviour in novel 2D semiconducting MXene $\bf{Ti_2CO_2}$}

%\title{Investigations on high-frequency transport in the semiconducting two-dimensional MXene $\bf{Ti_2CO_2}$ {\ensuremath{\mathrm{\bf{Ti_2CO_2}}}}}

\title{High-frequency complex impedance analysis of the two-dimensional semiconducting MXene-{\ensuremath{\bf{Ti_2CO_2}}}}

	\author{Anup Kumar Mandia$^{a}$}
	\author{Rohit Kumar$^{a}$}	
	\author{Namitha Anna Koshi$^{b}$}
	\author{Seung-Cheol Lee$^{c}$}
	%\thanks{corresponding author: leesc@kist.re.kr}
	\author{Satadeep Bhattacharjee$^{b}$}
    \author{Bhaskaran Muralidharan$^{a}$}
	\thanks{corresponding author: bm@ee.iitb.ac.in}
	\affiliation{$^{a}$Department of Electrical Engineering, Indian Institute of Technology Bombay, Powai, Mumbai-400076, India}
	\affiliation{$^{b}$Indo-Korea Science and Technology Center (IKST), Jakkur, Bengaluru 560065, India}
    \affiliation{$^{c}$Electronic Materials Research Center, KIST, Seoul 136-791, South Korea.}
\date{\today}

\begin{abstract}
%In this work, we compute the real and imaginary components of the electron mobility and conductivity of the two-dimensional (2D) semiconducting MXene Ti\textsubscript{2}CO\textsubscript{2} for high-frequency applications in relation to temperature and doping concentration. The calculations are carried out using our fully $ab~initio$ model developed for AC electric field. Rode's iterative method in conjunction with density functional theory (DFT) has been used in this model. The model is fully $ab~initio$ which requires inputs calculated from the first principle using DFT. We compare our results with those obtained from Drude's method to validate our model. We show that at lower frequencies, Drude's values are close to exact values and that mobility and conductivity can be estimated using Drude's model unless high accuracy is required whereas, at higher frequencies, the predicted mobility and conductivity values differ significantly from those obtained using Drude's theory. Hence, the results obtained from Drude's theory are not accurate enough for a meaningful comparison with experiments. In our calculations, the effect of acoustic deformation potential scattering, piezoelectric scattering, and polar optical phonon scattering mechanisms have been shown, which is significant in this material and was not highlighted earlier. The design and use of 2D semiconducting MXene Ti\textsubscript{2}CO\textsubscript{2} based devices are anticipated to be promising based on the current results.

\noindent The two-dimensional compound group of MXenes, which exhibit unique optical, electrical, chemical, and mechanical properties, are an exceptional class of transition metal carbides and nitrides.
In addition to traditional applications in Li-S, Li-ion batteries, conductive electrodes, hydrogen storage, and fuel cells, the low lattice thermal conductivity coupled with high electron mobility in the semiconducting oxygen-functionalized MXene  Ti\textsubscript{2}CO\textsubscript{2} has led to the recent interests in high-performance thermoelectric and nanoelectronic devices. Apart from the above dc- transport applications, it is crucial to also understand ac- transport across them, given the growing interest in applications surrounding wireless communications and transparent conductors. 
In this work, we investigate using our recently developed $ab~initio$ transport model, the real and imaginary components of electron mobility and conductivity to conclusively depict carrier transport beyond the room temperature for frequency ranges upto the terahertz range. We also contrast the carrier mobility and conductivity with respect to the Drude's model to depict its inaccuracies for a meaningful comparison with experiments. Our calculations show the effect of acoustic deformation potential scattering, piezoelectric
scattering, and polar optical phonon scattering mechanisms. Without relying on experimental data, our model requires inputs calculated from first principles using density functional theory. Our results set the stage for providing ab-initio based ac- transport calculations given the current research on MXenes for high frequency applications. %Our results illustrate the potential for Ti\textsubscript{2}CO\textsubscript{2} as active materials in nanoelectronic and optoelectronic devices, and hence the design and use of 2D semiconducting MXene Ti\textsubscript{2}CO\textsubscript{2}-based devices are anticipated to be promising based on the current results. 

\end{abstract}

\maketitle
\section{Introduction}

Two-dimensional (2D) materials \cite{mas20112d,novoselov2004electric,geim2009graphene,geim2010rise,neto2009electronic,novoselov2005two,akinwande2017review} have unique optical, electrical, chemical, and mechanical properties and have received much attention over the last two decades due to their utility in optoelectronics \cite{cheng2019recent}, flexible electronics \cite{kim2015materials}, nanogenerators \cite{wang2006piezoelectric}, nanoelectromechanical systems \cite{lee2018electrically}, sensing \cite{cai2018stretchable} and terahertz (THz) frequency devices \cite{lin2013terahertz}. The 2D compound group of MXenes \cite{barsoum2000mn+,barsoum2011elastic,naguib2011two,naguib2012two,naguib201425th,ghidiu2014synthesis}
%, which exhibit unique optical, electrical, chemical, and mechanical properties, 
are an exceptional class of transition metal carbides and nitrides. In addition to traditional applications in Li-S \cite{liang2015sulfur}, Li-ion batteries (LIBs) \cite{naguib2012mxene, tian2019flexible, xie2014role,tang2012mxenes,liang2015sulfur,er2014ti3c2}, conductive electrodes, \cite{halim2014transparent},
hydrogen storage \cite{hu2013mxene,kumar2021mxenes,lei2015recent,li2018functional,hu2014two}, optics \cite{wang2020mxenes,jiang2020two,fu2021mxenes,xue2017photoluminescent}, and fuel cells \cite{xie2013extraordinarily}, the narrow bandgap, improved thermodynamic stability, high Seebeck coefficient, high figure of merit, and the low lattice thermal conductivity coupled with high electron mobility in semiconducting and oxygen-functionalized 2D MXene  Ti\textsubscript{2}CO\textsubscript{2} has led to the recent interests in high-performance thermoelectric (TE) \cite{gandi2016thermoelectric,zha2016thermal,sarikurt2018influence} and nanoelectronic devices \cite{kim2019mxetronics,sarycheva20182d}. In the past few years, different modules \cite{madsen2006boltztrap, ponce2016epw, faghaninia2015ab, ganose2021efficient} have been developed to predict dc transport properties of such materials with inputs calculated from the first-principles methods. \\ 
\indent The importance of high-frequency analysis is growing due to the applicability of MXenes in developing thin and flexible antennas and for wearable and transparent electronic devices \cite{sindhu2022mxene,lee2021polymer,ma2021flexible,zhang2019graphene,kim2019mxetronics}.
%The importance of high frequency analysis is growing as a result of the applicability of MXenes in the development of thin and flexible antennas, and for wearable and transparent electronic devices 
There are several aspects that function as restrictions, which makes it more challenging to fabricate such an antenna for wireless communication. The skin depth is one of them, which is frequency dependent. %Therefore, to ensure a sufficient electrical current flow, the thicknesses of such MXene based antennas should be at least three to four times higher than their skin depth, hence this requires precise skin depth calculations. 
Apart from this, impedance matching is necessary, which further helps in determining the voltage standing wave ratio (VSWR). Typically, MXenes can match impedance at extremely small antenna thicknesses \cite{sarycheva20182d,pozar2009microwave}. Furthermore, the gain value depends on the thickness, and the gain values of all the MXene antennas decreases with decreasing thickness, which further emphasizes on accurate computation of frequency-dependent transport parameters. With the thickness and electrical conductivity, contributing majorly towards absorption, it is also possible to estimate the absorption loss for non-magnetic and conducting shielding materials \cite{iqbal20202d}.\\
\indent Given the myriad possibilities for high frequency operation of MXenes, and the paucity in the development of theoretical models, we advance an extended version of our previously developed module AMMCR \cite{mandia2021ammcr} to obtain the highly accurate frequency-dependent electrical conductivity and other high frequency transport characteristics, which paves the way for the manufacturing of a variety of flexible, wearable, and nanoelectronic devices.
%\indent  to calculate the mobility and the conductivity of 2D semiconductor materials for AC electric field. 
%This is the first step towards the development of a fully $ab~initio$ based carrier transport model for calculating the transport coefficients of 2D semiconducting materials for ac electric field.\\
\indent Our model is based on the Rode's iterative method \cite{rode1970electron,rode1973theory,rode1975low} evolved from the conventional semi-classical Boltzmann transport equation (BTE) \cite{lundstrom2002fundamentals}. We have included various scattering mechanisms explicitly. The variations in the characteristics across a wide range of temperatures, carrier concentrations, and frequencies are admirably captured by our model. Since our model does not rely on experimental data, it can predict the transport parameters of any new emerging semiconducting material that is currently being developed and for which the experimental data is unavailable. The potential capabilities of our model \cite{mandia2021ammcr,mandia2019ab,mandia2022electrical,kumar2023advancing} include lower computational time, less complexity, more precision \& accuracy, minimal computational expense, independent of experimental data, user-friendly, and faster convergence. \\ 
\indent First, we will outline the Rode's iterative methodology used in this study to evaluate the electron transport coefficients while taking into consideration the various scattering mechanisms at play in this semiconducting 2D material for high-frequency applications. Then, we will discuss the findings about Ti\textsubscript{2}CO\textsubscript{2} MXene's electron mobility and conductivity for ac electric field. 
\indent In the upcoming sections, we will outline Rode's iterative methodology developed here to evaluate the electron transport coefficients while taking into consideration the various scattering mechanisms for high-frequency applications. Then, we will discuss the findings about Ti\textsubscript{2}CO\textsubscript{2} MXene's electron mobility and conductivity for AC electric fields. All necessary $ab~initio$ inputs required for simulations are calculated from first-principles density functional theory (DFT) using the Vienna ab-initio Simulation Package (VASP) module without using any experimental data. Further, we investigate the temperature, carrier concentration and frequency dependence of mobility and conductivity in Ti\textsubscript{2}CO\textsubscript{2}. 
\section{Methodology}
\label{sec_methodology}
\subsection{Calculation of the \textit{ab initio} parameters}
We use density functional theory (DFT) to calculate the $\emph{ab~initio}$ parameters, and DFT implemented in the plane wave code Vienna ab-initio Simulation Package (VASP) \cite{kresse1996efficiency, kresse1996efficient} is used to carry out the electronic structure calculations. The projector augmented wave (PAW) \cite{blochl1994projector, kresse1999ultrasoft} method is used to calculate the pseudo-potentials, and the generalized gradient approximation (GGA) is used to treat the exchange-correlation functional, which is parameterized by the Perdew-Burke-Ernzerhof (PBE) formalism \cite{perdew1996generalized}. The cut-off energy for plane waves is set at $500 \:eV$, and for the structural optimization, the conjugate gradient algorithm is used. The energy and force convergence criteria are $10^{-6} \: eV$ and $-0.01\: eV/$\AA,\:respectively. The parameters needed for simulations are shown in table \ref{table1}. For more details on band structure and parameters needed for simulations, DOS and phonon dispersion, refer to our previous study \cite{mandia2022electrical}.

\begin{table}
	\centering
	\caption{  Material Parameters used for Ti\textsubscript{2}CO\textsubscript{2}}
	%\begin{ruledtabular}
	\begin{tabular}{|l|l|}
		\hline
		\bf{Parameters}  &   \bf{Values}  \\
		\hline
	PZ constant, $e_{11} $ (C/m)      &   $3 \times 10^{-13}$ \\
	 \hline
    Acoustic deformation potentials, $D_{A} (ev):$       &    \\
	$D_{A,LA} $  & 8.6     \\
    $D_{A,TA} $  & 3.5     \\
    $D_{A,ZA} $  & 0.7     \\
     \hline
    Elastic modulus, $C_{A} (N/m) :$ & 		\\
    $C_{A,LA}  $    & $301.7  $      \\
    $C_{A,TA}  $    & $391.6 $      \\
    $C_{A,ZA}  $    & $59.3 $      \\
     \hline
    Polar optical phonon frequency $\omega_{_{POP}} (THz):$ &  \\
    $\omega_{_{POP,LO}} $  & 3.89     \\
    $\omega_{_{POP,TO}} $  & 3.89     \\
    $\omega_{_{POP,HP}} $  & 8.51     \\
     \hline
    High frequency dielectric constant, $ \kappa_{_{\infty}}$  & 23.57 \\
     \hline
    Low frequency dielectric constant, $ \kappa_{_{0}}$     &    23.6 \\
    \hline
	\end{tabular}
	%\end{ruledtabular}
	\label{table1}
   \end{table}

\subsection{Solution of the Boltzmann transport equation}
Our model, which is a more evolved and advanced version of our previously developed tool AMMCR, is used to compute the transport parameters \cite{mandia2021ammcr,mandia2019ab}.
The brief methodology for solving the BTE is presented below. The BTE for the electron distribution function $f$ is given by

\begin{equation}
\frac{\partial f(\textbf{k})}{\partial t} + \textbf{v}\cdot\nabla _rf - \frac{e \textbf{F}}{\hbar} \cdot\nabla _kf= \frac{\partial f}{\partial t}\Bigr|_{\substack{coll}},
\label{BTE}
\end{equation}

\noindent where $\textbf{v}$ is the carrier velocity, $e$ is the electronic charge, $\textbf{F}$ is the applied electric field, and $f$ represents the probability distribution function of carrier in the real and the momentum space as a function of time, $\frac{\partial f}{\partial t}\Bigr|_{\substack{coll}}$ denotes the change in the distribution function with time due to the collisions. Under steady-state, $\frac{\partial f(\textbf{k})}{\partial t} = 0$ and spatial homogeneous condition $\nabla_rf=0$, Eq. (\ref{BTE}) can be written as 

\begin{equation}
\begin{split}
\frac{-e {\textbf{F}}}{\hbar}\cdot\nabla _kf = \int \Big[ s(\textbf{k$^\prime$, k})\; f'(1-f)\\
-  s(\textbf{k, k$^\prime$})f(1-f')\Big] d\textbf{k$^\prime$},
\end{split}
\label{BTE1}
\end{equation}

\noindent where $s(\textbf{k}, \textbf{k}^\prime)$ denotes the transition rate of an electron from a state $\textbf{k}$ to a state $\textbf{k}'$. For better understanding of the Rode's algorithm \cite{rode1970electron,Rode_II-IVSC,rode1975low,rode1973theory}, refer to our previous studies \cite{mandia2021ammcr,mandia2019ab}. 

\subsection{Complex mobility and conductivity}

For an ac electric field  ${\textbf{F} = F_{0}}$\:$e^{j \omega t}$, the distribution function can be expressed as \cite{ghosal1982high,nag2012electron,nag1975galvanomagnetic}

\begin{equation}
f(\textbf{k}) = f_{0}(k) - {e v F} \Big[\phi_{r} + j \phi_{i}\Big]\:\Big(\frac{\partial f_{0}}{\partial E}\Big) \: cos\: \theta_c\: ,
\label{pert_ac}
\end{equation}

\noindent where $\textbf{k}$ is the Bloch vector for energy $E$,  $\theta_c$ is the angle between $\textbf{F}$ and $\textbf{k}$. Here, $k$=$|\textbf {k}|$, $F$=$|\textbf {F}|$, and $\phi_{r}$ and $\phi_{i}$ are functions that can be calculated using the Boltzmann equation. Substituting Eq. (\ref{pert_ac}) in the BTE and equating the coefficient of $\phi_{r}$ and $\phi_{i}$ on two sides, we get 

\begin{equation}
L_{c}\phi_{r} = 1 + \phi_{i} \omega 
\label{Lc1}
\end{equation}

\begin{equation}
L_{c}\phi_{i} + \omega \phi_{r} = 0\:,  
\label{Lc2}
\end{equation}

\noindent where $L_c$ is the collision operator, and it is given by \cite{nag1975microwave,howarth1953theory} 

\begin{equation}
\begin{split}
L_{c}\:\phi(E) = S_o(E)\: \phi(E) - S_e(E)\: \phi(E + \hbar\: \omega_{pop})\\
- S_a(E)\: \phi(E - \hbar\: \omega_{pop}) \:, 
\end{split}
\label{Lc3}
\end{equation}

\noindent where $\hbar\: \omega_{pop}$ is the polar optical phonon (POP) energy. The term $S_o(E)$ is the sum the of out-scattering rates due to the POP scattering interaction and the out-scattering and in-scattering contributions from the elastic scattering processes. Here, $S_e(E)$ and $S_a(E)$ are the in-scattering rates due to the emission and the absorption processes of the POP scattering mechanism. 
Separating $\phi_{r}$ and $\phi_{i}$ from (\ref{Lc1}) and (\ref{Lc2}), we get

\begin{equation}
\begin{split}
\phi_{r}(E) = \frac{S_o(E)}{S_o^{^2}(E) + \omega^2}  \left[ 1 + S_{ar}(E) + S_{er}(E)\right] \\
+ \frac{\omega}{S_o^{^2}(E) + \omega^2} \left[S_{ai}(E) + S_{ei}(E)\right] 
\end{split}
\label{fi_real}
\end{equation}

\begin{equation} 
\begin{split}
\phi_{i}(E) = \frac{-\omega}{S_o^{^2}(E) + \omega^2}  \left[ 1 + S_{ar}(E) + S_{er}(E)\right] \\
+ \frac{S_o(E)}{S_o^2(E) + \omega^2} \left[S_{ai}(E) + S_{ei}(E)\right] \:,
\end{split}
\label{fi_img}
\end{equation}

\noindent where $S_{ar}(E) = S_{a}(E) \ \phi_r(E - \hbar\: \omega_{pop})$, $S_{er}(E) = S_{e}(E) \ \phi_r(E + \hbar\: \omega_{pop})$, $S_{ai}(E) = S_{a}(E) \ \phi_i(E - \hbar\: \omega_{pop})$ and $S_{ei}(E) = S_{e}(E) \ \phi_i(E + \hbar\: \omega_{pop})$. Equations (\ref{fi_real}) and (\ref{fi_img}) are to be solved by the Rode's iterative method \cite{rode1973theory, rode1975low}.\\
After determining $\phi_{r}$ and $\phi_{i}$, the real and imaginary components of complex conductivity are determined using the following expressions \cite{nag1975galvanomagnetic,nag1975microwave}

\begin{equation}
\sigma_{r} =   \frac{e^2 \int v^2(E) D_s(E) \phi_{r}(E) (\frac{\partial f_{0}}{\partial E}) dE}{2F} 
\label{sigma_real}
\end{equation}

\begin{equation}
\sigma_{i} =   \frac{e^2 \int v^2(E) D_s(E) \phi_{i}(E) (\frac{\partial f_{0}}{\partial E}) dE}{2 F} \:,
\label{sigma_img}
\end{equation}

\noindent where $D_s(E)$ is the density of states (DOS).
The complex mobility can be expressed as $\mu_{r}-j\mu_{i}$. The real and imaginary components of mobility are then calculated by using the following equation

\begin{equation}
\mu_r= \frac{\sigma_r}{N_{_D} e}\times t_{_Z}
\label{mu_r}
\end{equation}

\begin{equation}
\mu_i= \frac{\sigma_i}{N_{_D} e}\times t_{_Z}\:,
\label{mu_i}
\end{equation}

\noindent where $N_{_D}$ is the electron doping concentration and $t_{_z}$ denotes the thickness of the material along the z-direction, which is taken to be 4.45 \AA \: in our calculations.

\subsection{Scattering Mechanisms}
\subsubsection*{\bf\emph{1. \hspace{0.2cm}{Acoustic deformation potential scattering}}}
The scattering rate of the acoustic deformation potential (ADP) scattering is defined as \cite{zha2016thermal,hwang2008acoustic,lin2014terahertz}
\begin{equation}
\frac{1}{\tau_{_{ADP}}^\kappa(E)} = \frac{D_{\kappa A}^2 k_{B} T k}{\hbar^2 C_{A}^\kappa v} \:,
\label{acoustic_rate}
\end{equation}

\noindent where $k$ is the wave vector, $ C_{A} $ is the elastic modulus, $T$ is the temperature, $ \hbar $ is the reduced Planck's constant, $D_{\kappa A}$ is acoustic deformation potential and $ k_{B} $ denotes the Boltzmann constant. The term $v$ is the group velocity of the electrons, and $\kappa \in LA, TA, ZA$. The energy dependence is introduced here, via the wave vector $k$. We fit the lowest conduction band analytically with a six-degree polynomial to obtain a smooth curve for the group velocity. This allows us to obtain a one-to-one mapping between the wave vector $k$ and the band energies.

\subsubsection*{\bf\emph{2. \hspace{0.2cm}{Piezoelectric scattering}}}
The scattering rates due to piezoelectric (PZ) scattering can be expressed as \cite{kaasbjerg2013acoustic}

\begin{equation}
\frac{1}{\tau_{_{PZ}}(E)} = \frac{1}{\tau_{_{ADP}}(E)} \times \frac{1}{2}\times \left( \frac{e_{11} e}{\epsilon_0 D_{A}}\right) ^2 \:,
\label{Piezo_rate}
\end{equation}

\noindent where $\epsilon_0$ is vacuum permeability, and $e_{11}$ is a piezoelectric constant (unit of C/m).   

\begin{figure}[!t]
	\centering
	{\includegraphics[height=0.35\textwidth,width=0.45\textwidth]{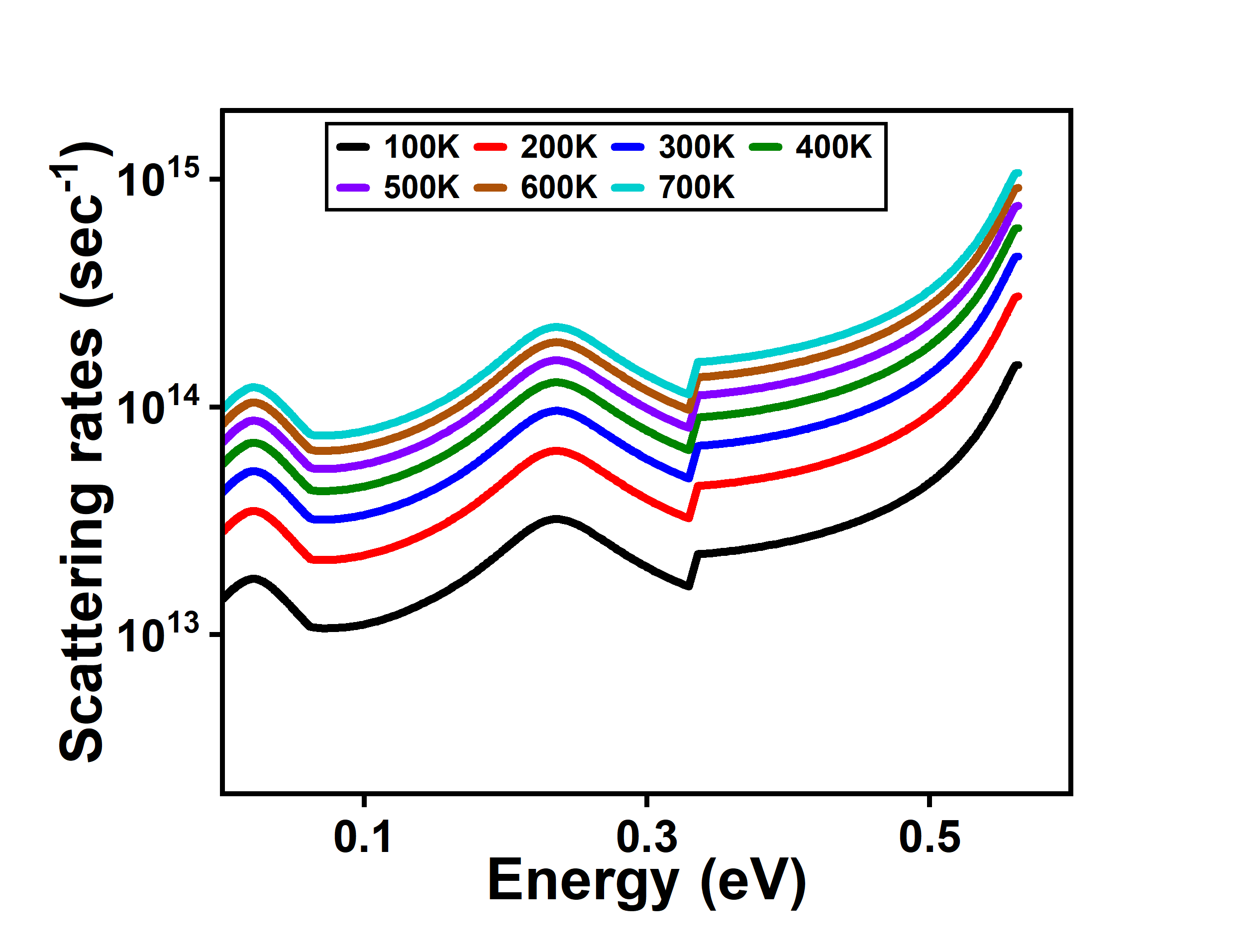}}
	\quad
	\caption{Scattering rates vs. carrier energy at various temperatures with a doping concentration of $1\times 10^{12}\: cm^{-2}$.}
	\label{Scattering rates}
\end{figure}

\begin{figure}[!t]
	\centering
	{\includegraphics[height=0.35\textwidth,width=0.45\textwidth]{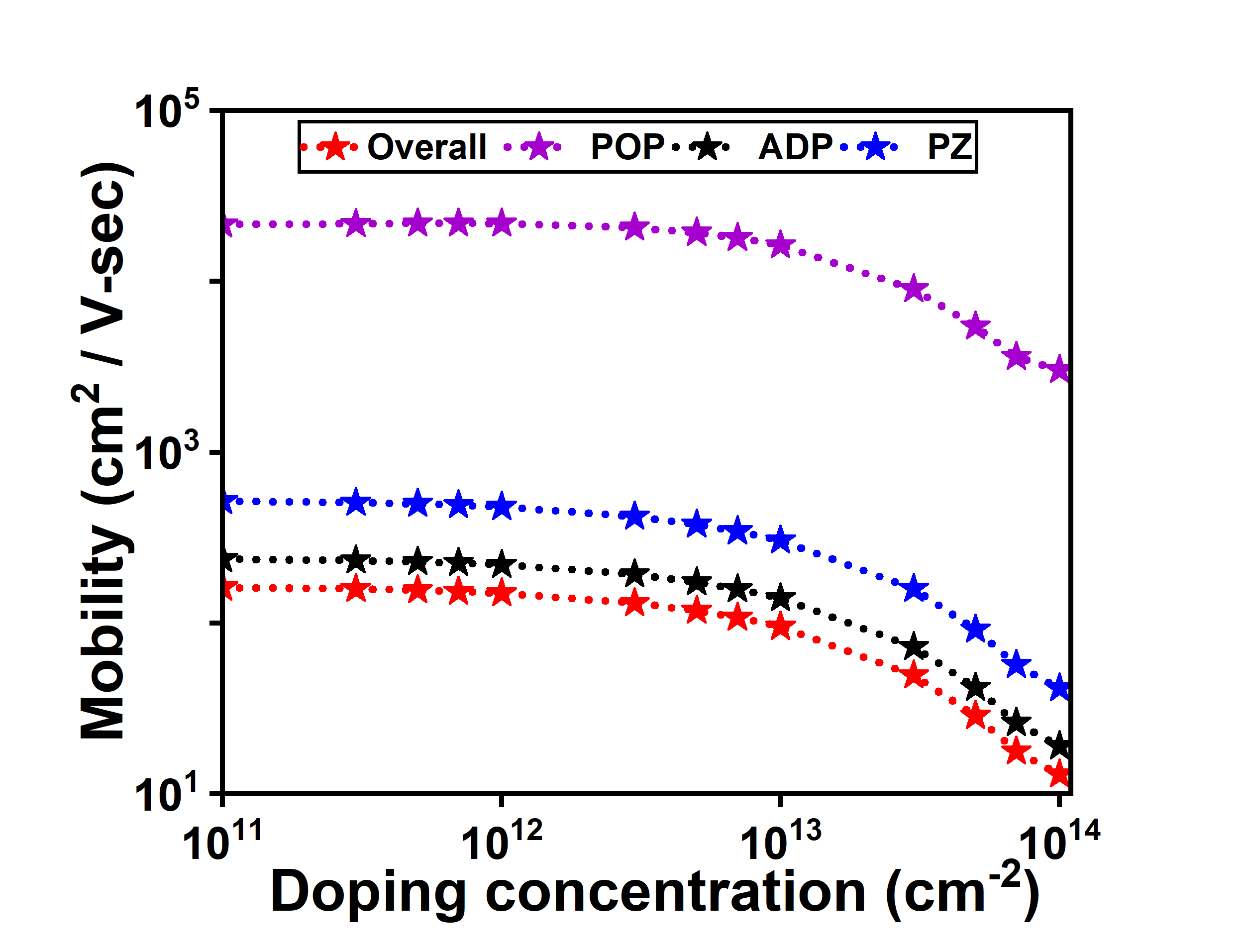}}
	\quad
	\caption{Contribution to the mobility by different scattering mechanisms vs. doping concentration at room temperature with zero Hz frequency.}
	\label{Mobility contribution}
\end{figure}

\subsubsection*{\bf\emph{3. \hspace{0.2cm}{Polar optical phonon scattering}}}
The inelastic scattering in the system is caused by POP scattering mechanism. The POP scattering contribution to the out-scattering is given by \cite{nag2012electron,kawamura1992phonon}

\begin{equation}
\begin{split}
S_{o}^{in}(k) = \frac{C_{_{POP}}}{(1- f_{0}(E))} [P+Q]\:,
\end{split}
\label{out_sc_pop}
\end{equation}
where 
\begin{equation}
P=N_V (1 - f_{0}(E + \hbar \omega_{_{POP}}) I^+(E) \frac{k^+}{v(E + \hbar \omega_{_{POP}})}
\label{P}
\end{equation}

%\begin{widetext}
\begin{equation}
\begin{split}
Q=(N_V+1) (1 - f_{0}(E - \hbar \omega_{_{POP}}) \\
\times ~ I^-(E) \frac{k^-}{v(E - \hbar \omega_{_{POP}})}
\end{split}
\label{Q}
\end{equation}
%\end{widetext}

\begin{equation}
I^{+}(E) = \int_0^{2\pi} \frac{1}{q_{a}} d\theta 
\label{I_plus}
\end{equation}

\begin{equation}
I^{-}(E) = \int_0^{2\pi} \frac{1}{q_{e}} d\theta 
\label{I_minus}
\end{equation}

\begin{equation}
q_{a} = \left( k^2 + \left( k^{+}\right)^2 - 2k k^{+} cos\:\theta \right)
\label{q_a}
\end{equation}

\begin{equation}
q_{e} = \left( k^2 + \left( k^{-}\right)^2 - 2k k^{-} cos\:\theta \right)\:,
\label{q_e}
\end{equation}

\noindent where $k^+$ denotes the wave vector at energy $E+ \hbar\: \omega$ and  $k^-$ represents the wave vector at energy $E- \hbar\: \omega$. The angle between the initial wave vector $\textbf{k}$ and the final wave vector $\textbf{k}^{'}$ is defined as $\theta $.

\begin{equation}
C_{_{POP}} = \frac{e^2 \omega_{_{POP}}}{8 \pi \hbar \epsilon_0} \times \left( \frac{1}{\kappa_{_{\infty}}} - \frac{1}{\kappa_{_0}} \right) \:,
\label{POP_const}
\end{equation}

\noindent where $\kappa_{_{\infty}}$ and $\kappa_{_{0}}$ are the high and low-frequency dielectric constants, respectively.
 
The sum of in-scattering rates due to the POP absorption and the emission can be used to represent the in-scattering contribution due to the POP, an inelastic and anisotropic scattering mechanism.

\begin{equation}
S_i^{in}(k) = S_a^{in}(k) + S_e^{in}(k)\:,
\label{in_sc_pop}
\end{equation}

\noindent where $S_a^{in}(k)$ denotes the in-scattering of electrons from energy $E-\hbar\:\omega_{_{POP}}$ to energy $E$ due to the absorption of polar optical phonons and $S_e^{in}(k)$ denotes the in-scattering of electrons from energy $E+\hbar \omega_{_{POP}}$ to energy $E$ due to the emission of polar optical phonons.
   
\begin{equation}
S_{a}^{in}(k) = C_{_{POP}} (N_V+1) f_{0}(E)  J^-(E) \times A
\label{ab_in_sc_pop}
\end{equation}

\begin{equation}
A=\frac{k^-}{v(E- \hbar\: \omega_{_{POP}}) f_{0}(E - \hbar\: \omega_{_{POP}})}
\label{A}
\end{equation}

\begin{equation}
S_{e}^{in}(k) = C_{_{POP}} (N_V) f_{0}(E)  J^+(E) \times B 
\label{em_in_sc_pop}
\end{equation}

\begin{equation}
B=\frac{k^+}{v(E+ \hbar\: \omega{_{_{POP}}}) f_{0}(E + \hbar\: \omega_{_{_{POP}}})}
\label{B}
\end{equation}

\begin{equation}
N_V = \frac{1}{exp(\hbar\:\omega_{_{POP}}/k_B T) - 1}
\label{N}
\end{equation}

\begin{equation}
J^{+}(E) = \int_0^{2\pi} \frac{cos \theta}{q_{_{i,e}}}  d\theta 
\label{J_plus}
\end{equation}

\begin{equation}
J^{-}(E) = \int_0^{2\pi} \frac{cos \theta}{q_{_{i,a}}} d\theta 
\label{J_minus}
\end{equation}

\begin{equation}
q_{_{i,a}} = \Big[ \left( k^{-}\right)^2 + k^2 - 2k k^{-} cos\:\theta \Big]
\label{q_ab}
\end{equation}

\begin{equation}
q_{_{i,e}} = \Big[ \left( k^{+}\right)^2 + k^2  - 2k k^{+} cos \: \theta \Big]\:.
\label{q_em}
\end{equation}
\indent We replace the term $\frac{\hbar k}{m^*}$ by the group velocity \cite{mandia2021ammcr,mandia2019ab} while deriving the expression for different scattering rates, which we calculate using the DFT band structure \cite{mandia2019ab,mandia2021ammcr}.

\section{Results and Discussion}
\label{sec_result}

\begin{figure}[!t]
	\centering
	{\includegraphics[height=0.35\textwidth,width=0.45\textwidth]{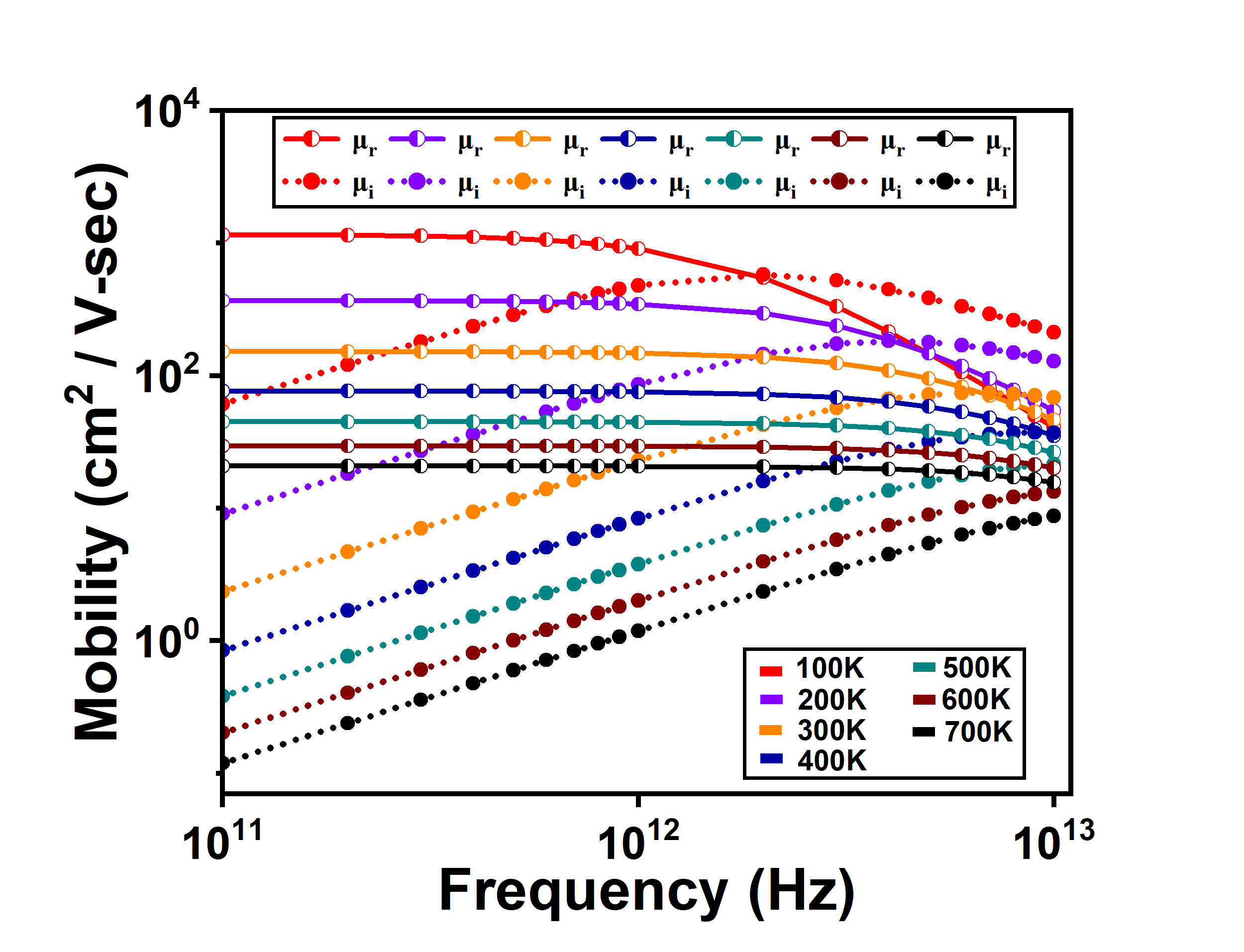}}
	\quad
	\caption{Variation of the real and the imaginary components of mobility as a function of frequency for various temperatures with a doping concentration of $1\times 10^{12}~ cm^{-2}$.}
	\label{Mob at various temp}
\end{figure}
In Fig. \ref{Scattering rates}, we show the total scattering rates, which is the sum of three scattering mechanisms we have considered in Ti\textsubscript{2}CO\textsubscript{2}. These three scattering mechanisms are the POP scattering, the PZ scattering, and the ADP scattering mechanisms. Scattering rates increase as we increase the temperature and hence the mobility decreases with an increase in the temperature.\\
%, which is explained in detail in the later paragraph.\\ 
\indent In Fig. \ref{Mobility contribution} we show the contribution to the mobility due to the ADP, the POP, and the PZ scattering mechanisms. 
%The acoustic phonon scattering, also known as 
Here, in Ti\textsubscript{2}CO\textsubscript{2}, the ADP scattering makes the most significant contribution followed by the PZ scattering mechanism and the POP scattering mechanism has the least effect. Therefore, the mobility and the conductivity in Ti\textsubscript{2}CO\textsubscript{2} is mainly limited by the ADP scattering and hence acoustic phonons are the primary limiters of the conductivity and the mobility. Refer to our previous study \cite{mandia2022electrical} where we have computed the scattering rates due to the individual acoustic phonons for better understanding the nature of acoustic phonons that limit the conductivity in Ti\textsubscript{2}CO\textsubscript{2}.\\
\indent In our previous study, we demonstrated that the longitudinal acoustic (LA) phonons play an important role in Ti\textsubscript{2}CO\textsubscript{2}. Therefore, the ADP and the PZ scattering are the two primary carrier scattering processes involved in Ti\textsubscript{2}CO\textsubscript{2}. As expected, the mobility decrease with increasing electron concentration. The mobility values are not significantly different for low electron concentrations, i.e., within the range [$1\times 10^{11}~cm^{-2} - 1\times 10^{13}~cm^{-2}$]. However, there is a sharp reduction in the mobility, as shown in Fig. \ref{Mobility contribution}, at electron concentrations greater than $1\times 10^{13}~cm^{-2}$. Consider Matthiessen's rule (\ref{matthiessen's rule}) to understand the mobility trend in relation to temperature and the carrier concentration, which can be expressed as 
\begin{figure}[!t]
	\centering
	{\includegraphics[height=0.35\textwidth,width=0.45\textwidth]{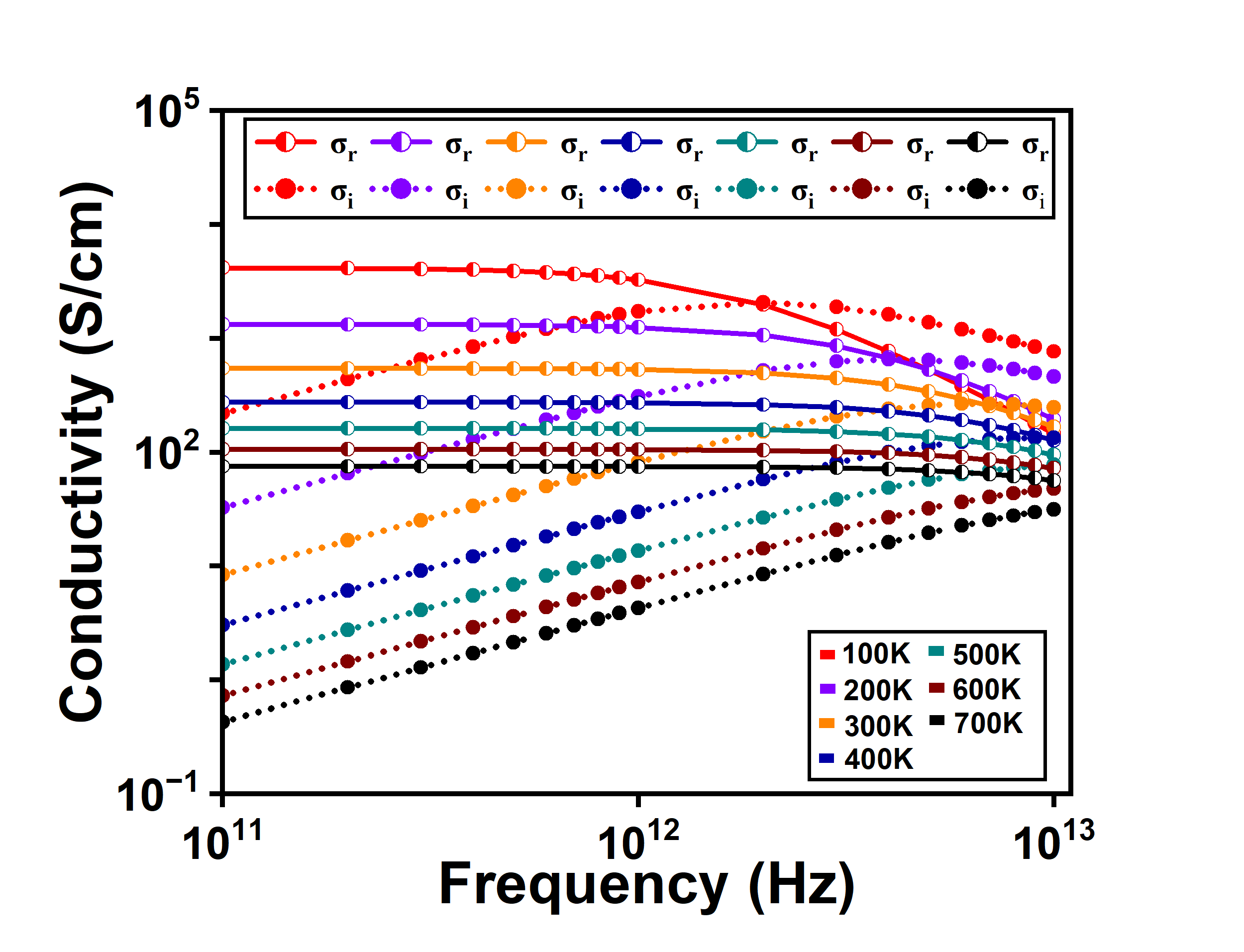}}
	\quad
	\caption{Variation of the real and the imaginary components of conductivity as a function of frequency for various temperatures with a doping concentration of $1\times 10^{12}~ cm^{-2}$.}
	\label{conductivity at various temp}
\end{figure}

\begin{equation}
	\frac{1}{\mu}=\frac{1}{\mu_{_{ADP}}}+ \frac{1}{\mu_{_{PZ}}}+\frac{1}{\mu_{_{POP}}}\:,
	\label{matthiessen's rule}
\end{equation}
\noindent where $\mu$ stands for total mobility.
%and the suffixes ADP, PZ and POP denote the acoustic deformation potential, piezoelectric and polar optical phonon contributions, respectively. 
Equation (\ref{matthiessen's rule}) states that the circumstance where the component with the lowest value is the most significant derives from the reciprocal relationship between the overall mobility and individual components. In Fig. \ref{Mobility contribution}, we show that the acoustic mode has the dominant contribution at all doping concentrations however, as demonstrated in Fig. \ref{Mobility contribution}, the PZ scattering mechanism significantly contributes across the entire doping range. 
\begin{figure}[!t]
	\centering
	{\includegraphics[height=0.35\textwidth,width=0.45\textwidth]{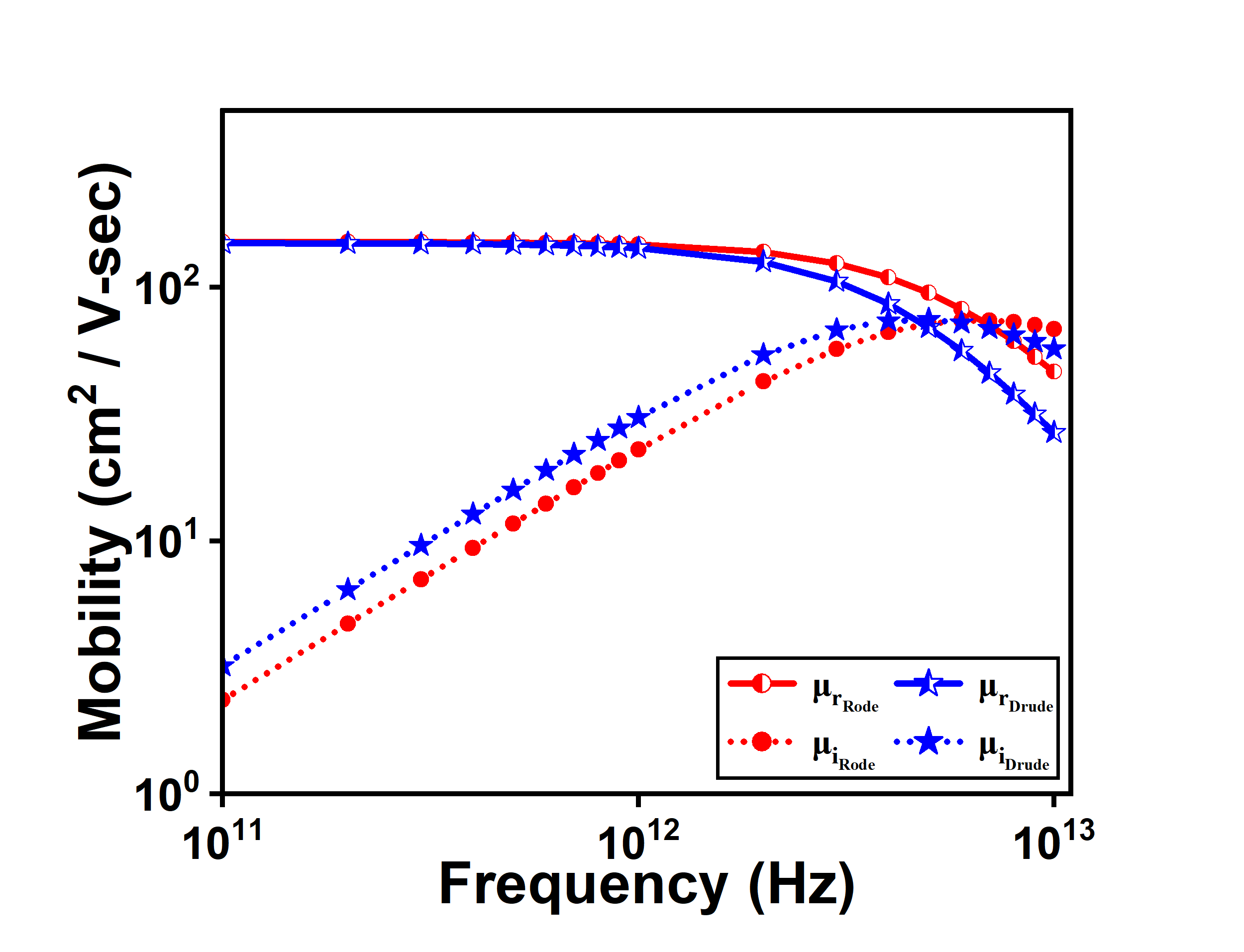}}
	\quad
	\caption{Comparison of mobility (i.e., the real and the imaginary components) with the Rode's and the Drude's method at 300 K for doping concentration of $1\times 10^{12}~cm^{-2}$.}
	\label{Mobility comparison}
\end{figure}

\begin{figure}[!t]
	\centering
	{\includegraphics[height=0.35\textwidth,width=0.45\textwidth]{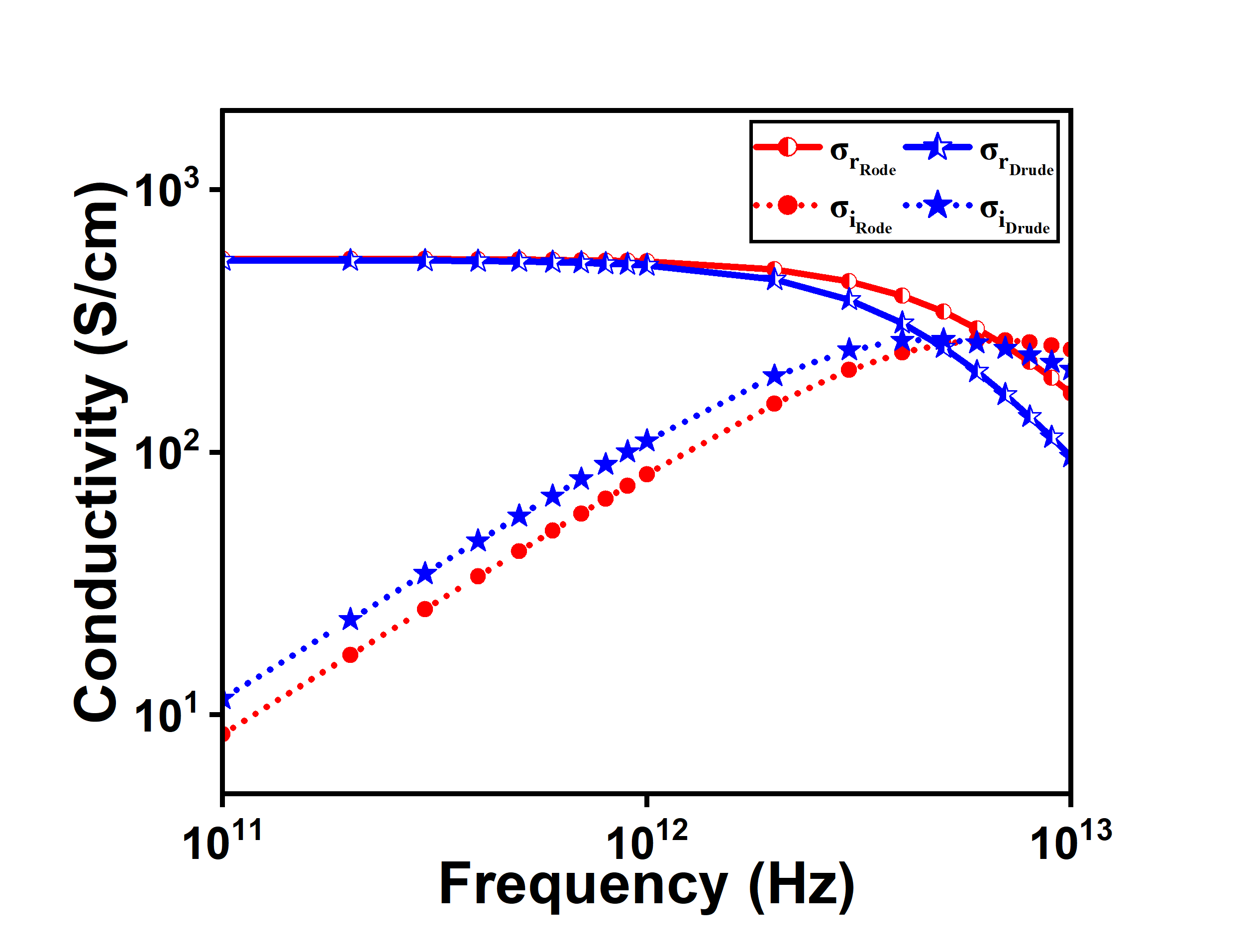}}
	\quad
	\caption{Comparison of conductivity (i.e., the real and the imaginary components) with the Rode's and the Drude's method at 300 K for doping concentration of $1\times 10^{12}~cm^{-2}$.}
	\label{Conductivity comparison}
\end{figure}
In Fig. \ref{Mob at various temp}, we show that as the temperature rises from 100 K to 700 K, both $\mu_r$ and $\mu_i$ for $N_D=1\times 10^{12}~cm^{-2}$ decreases and these findings are explained by the fact that the scattering rate increases with increasing temperature. Figure \ref{Mob at various temp} depicts the crossover between $\mu_r$ and $\mu_i$ at a specific frequency for each temperature, with the crossover shifting towards a higher frequency as the temperature increases. It is worth pointing that high-frequency effects are stronger at temperatures below the room temperature. Figure \ref{conductivity at various temp} shows the decrease in the conductivity caused by a reduction in mobility with an increase in temperature.
%, as we have shown above. 
In Fig. \ref{Mobility comparison}, we compare the mobility calculated using the Rode's iterative method to the mobility calculated using the Drude's theory. The Drude's mobility can be expressed as \cite{chattopadhyay1980temperature}

\begin{equation}
\mu_r=\frac{(e\tau/m^*)}{(1+\omega^2 \tau^2)} 
\label{Drude_real_mob}
\end{equation}

\begin{equation}
\mu_i=\frac{(e \omega \tau^2/m^*)}{(1+\omega^2 \tau^2)} 
\label{Drude_imag_mob}
\end{equation}

\begin{figure}[!t]
	\centering
	{\includegraphics[height=0.35\textwidth,width=0.45\textwidth]{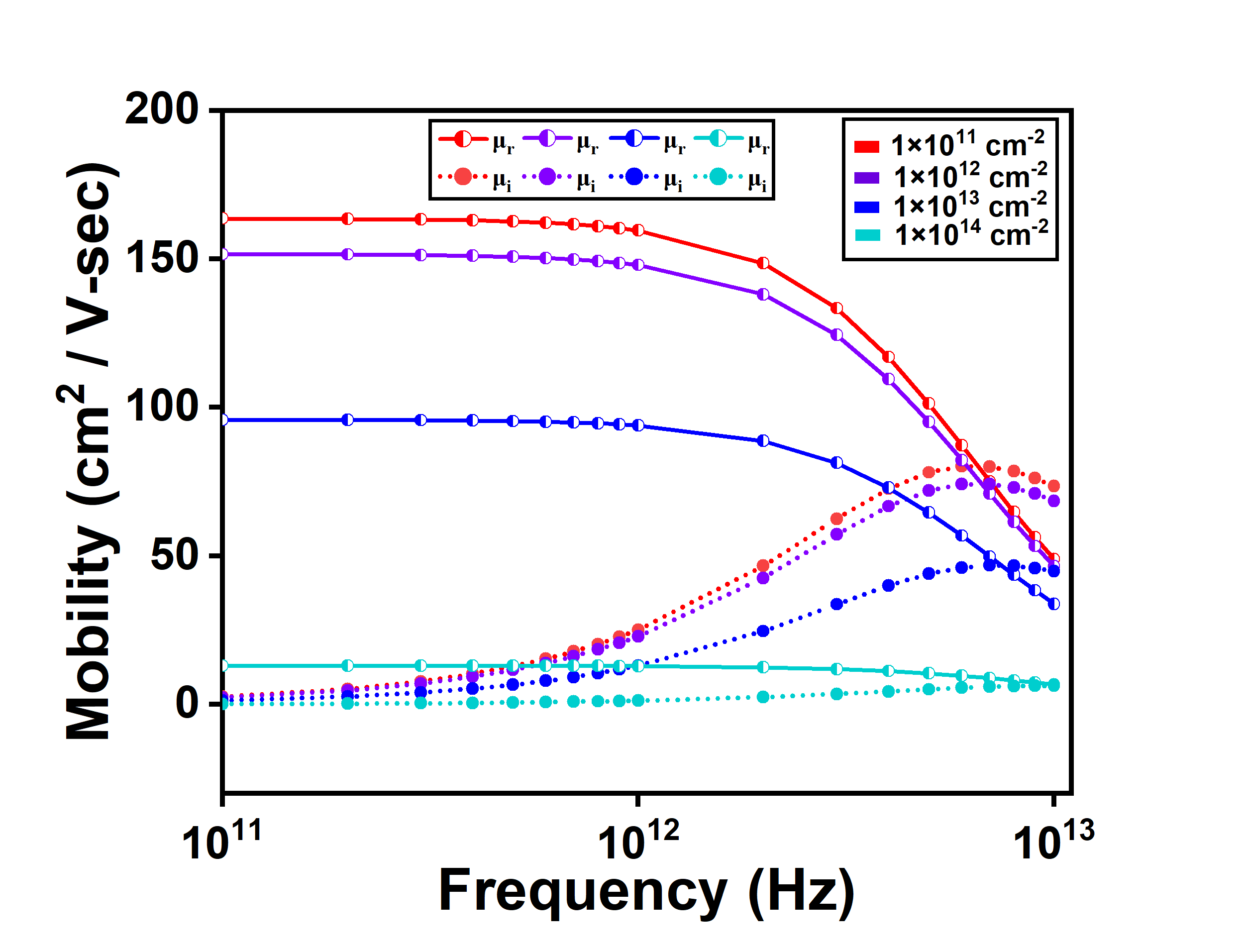}}
	\quad
	\caption{Variation of the real and the imaginary components of the room temperature mobility as a function of frequency for various doping concentrations.}
	\label{Mob vs doping}
\end{figure}

\begin{figure}[!t]
	\centering
	{\includegraphics[height=0.35\textwidth,width=0.45\textwidth]{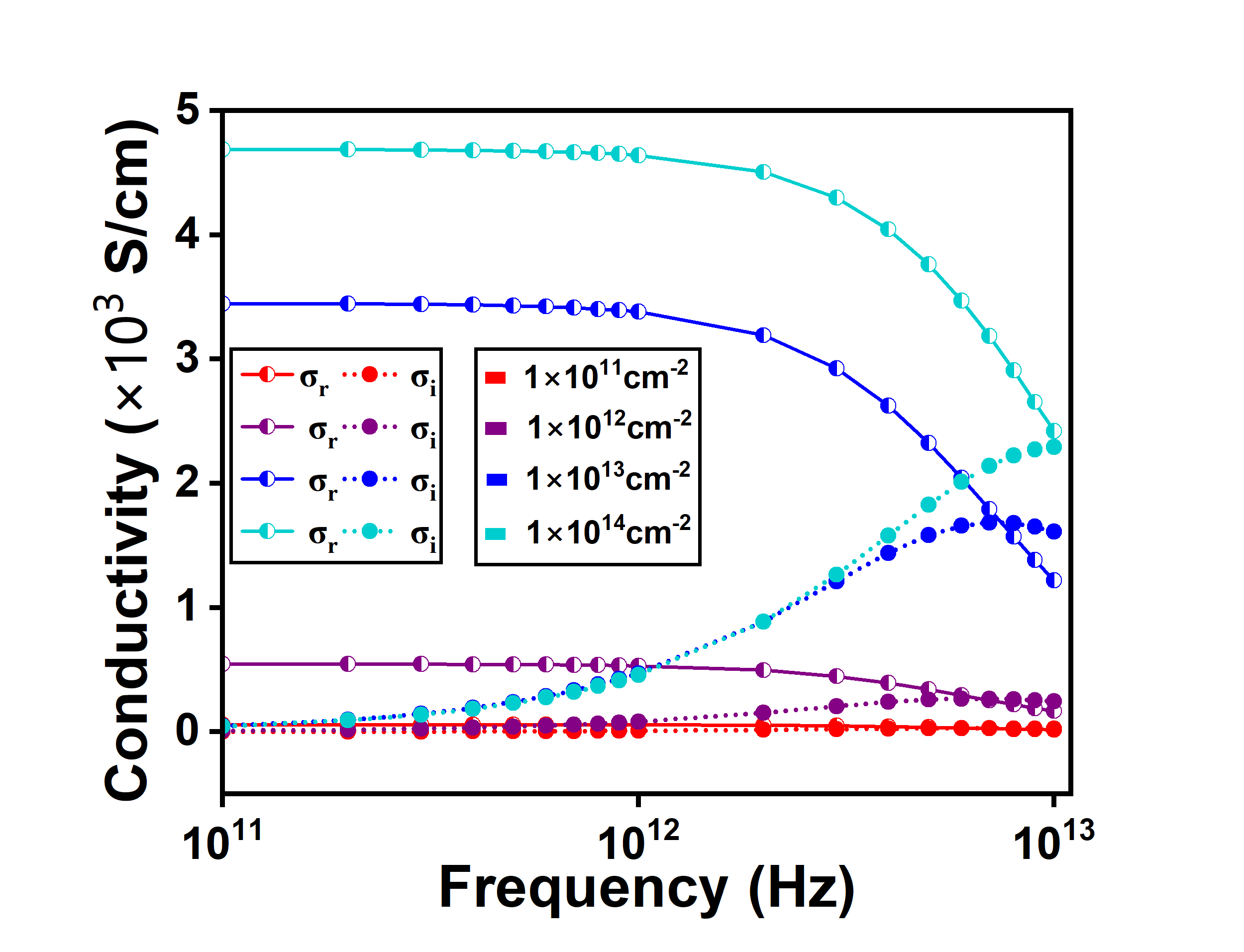}}
	\quad
	\caption{Variation of the real and the imaginary components of conductivity as a function of frequency for various doping concentrations at room temperature.}
	\label{Conductivity vs doping}
\end{figure}

Using the DFT, we calculate $m^*/m_0$, which is 0.4010, where $m_0$ is the rest mass of an electron. In order to determine the mobility using the Drude's approach, we first determine $\tau$ value at $\omega=0$ using $\mu_{r}$ as $1.496985 \times 10^{2}$ $cm^{2}/V-sec$, calculated at T =300 K and doping value of $1\times10^{12}~cm^{-2}$. The calculated $\tau$ value is $3.4134\times 10^{-14}$ seconds. By adjusting the value of $\omega$ while keeping $\tau$ fixed at $3.4134\times 10^{-14}$ seconds, we calculate the real and the imaginary components of mobility. 

\begin{figure}[!t]
	\centering
	{\includegraphics[height=0.35\textwidth,width=0.45\textwidth]{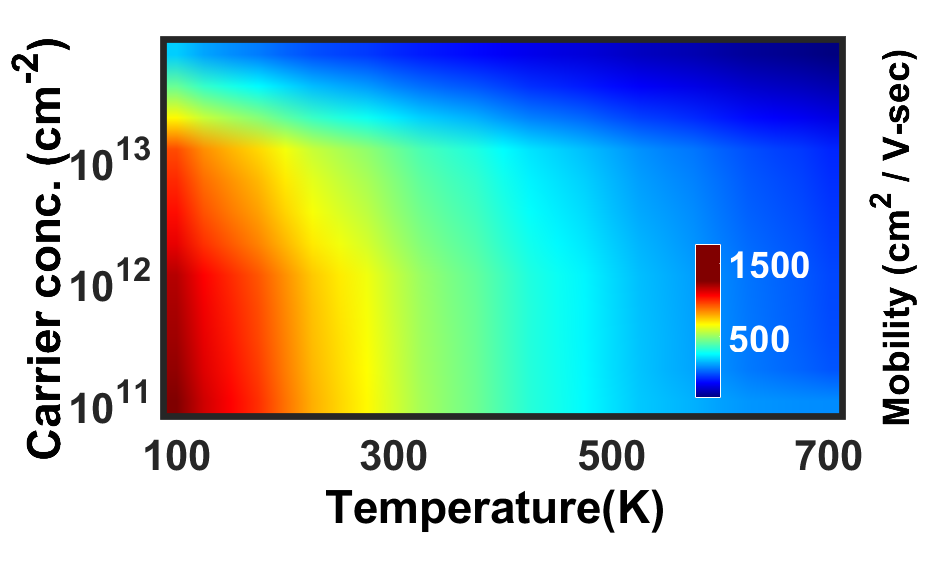}}
	\quad
	\caption{Mobility vs. temperature and carrier concentration at 100 GHz frequency.}
	\label{Mobility at 1e11 Hz}
\end{figure}

Figure \ref{Mobility comparison} shows that the Drude's values are close to exact values and that mobility can be estimated using the Drude's model unless high accuracy is required. But at higher frequencies, there is a significant deviation in the real component of mobility, and the Rode's method provides more accurate results.
%. Hence, it is better to use the Rode's approach at higher frequencies.
We also show the comparison of conductivity in Ti\textsubscript{2}CO\textsubscript{2} calculated using the Rode's and the Drude's approach in a manner similar to that of the mobility. The calculation of Drude's conductivity is straight forward using Eqs. (\ref{mu_r}) \& (\ref{mu_i}). As shown in Fig. \ref{Conductivity comparison}, Drude's values for the real conductivity are relatively close to the exact values for the conductivity and hence the Drude's method can be used to predict the mobility unless a high degree of accuracy is not required. At higher frequencies, a considerable variation has been seen, as a result the Rode's approach is more faithful at higher frequencies.\\
\indent In the Drude's model, the effect of all scattering mechanisms is merged together in a single constant relaxation time. This simplification makes all the calculations easy however, this method has some disadvantages (i) $\tau$ is obtained by fitting the experimental results to the mobility, which limits the predictability of this model. (ii) Due to over simplification of scattering mechanisms, it may lead to inaccurate results. (iii) It does not provide any insight into which scattering mechanisms are playing the major role. Instead, we have calculated all required inputs from the DFT without relying on any experimental data, included all scattering mechanisms explicitly and provided an insight regarding which scattering mechanisms are physically relevant to the considered semiconductor. Hence, our approach can be used to predict the mobility and the conductivity of even new emerging materials accurately. This paves the way for researchers in designing semiconductors with highly accurate transport properties based on theoretical predictions. \\
\begin{figure}[!t]
	\centering
	{\includegraphics[height=0.35\textwidth,width=0.45\textwidth]{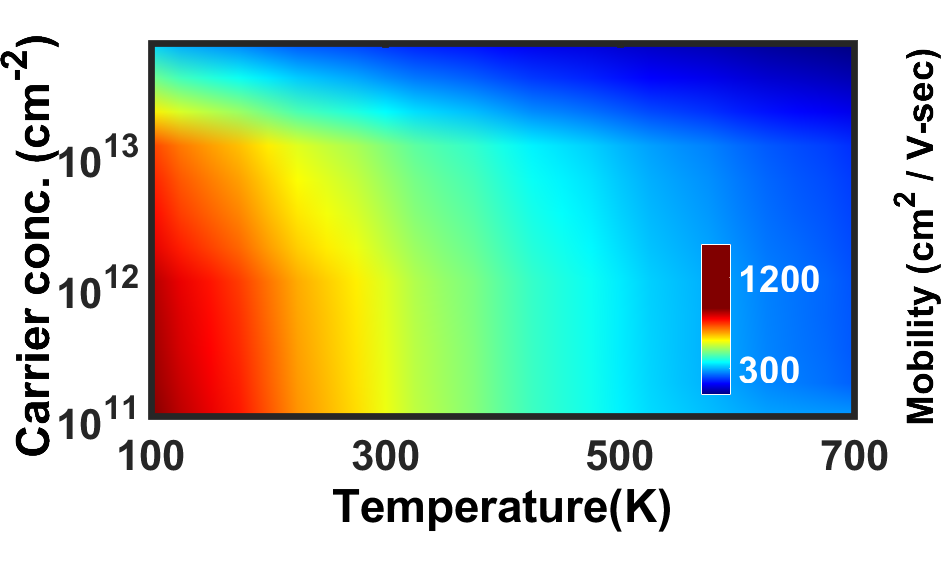}}
	\quad
	\caption{Mobility vs. temperature and carrier concentration at 1 THz frequency.}
	\label{Mobility at 1e12 Hz}
\end{figure}

Figure \ref{Mob vs doping} shows that  $\mu_r$ remains constant at room temperature up to a frequency of around 500 GHz. Beyond this frequency, $\mu_r$ falls while $\mu_i$ rises to a maximum and then falls. The simple Drude's expressions easily predict this type of behavior. Figure \ref{Mob vs doping} also shows that the mobility decreases as doping concentration increases. Figure \ref{Conductivity vs doping} depicts the increase in the conductivity caused by an increase in carrier concentration. For low values of frequency, $\sigma_r$ and $\sigma_i$ are nearly constant whereas at higher frequencies, above $10^{12}$ Hz, $\sigma_r$ decreases significantly while $\sigma_i$ increases. Figures \ref{Mobility at 1e11 Hz} and \ref{Mobility at 1e12 Hz} depict the mobility vs. temperature and carrier concentration at frequencies of $100$ GHz and $1$ THz, respectively. It is clear from both the figures that the mobility decreases as the frequency value increases and %It can also be seen in both figures that 
mobility increases as the carrier concentration and temperature decrease.
\section{Conclusion}
\label{conclu}
In this work, we investigated using our recently developed $ab~initio$ transport model, the real and imaginary components of electron mobility and conductivity to conclusively depict carrier transport beyond the room temperature for frequency ranges upto the terahertz range. We also contrasted the carrier mobility and conductivity with respect to the Drude model to depict its inaccuracies for a meaningful comparison with experiments. Our calculations showed the effect of acoustic deformation potential scattering, piezoelectric
scattering, and polar optical phonon scattering mechanisms. Without relying on experimental data, our model requires inputs calculated from first principles using density functional theory. Our results set the stage for providing ab-initio based ac- transport calculations given the current research on MXenes for high frequency applications.

\section*{Acknowledgments}
%\parskip{0pt}
%\setlength{\parskip}{0pt}
The authors gratefully acknowledge the funding from Indo-Korea Science and Technology Center (IKST), Bangalore. The author BM wishes to acknowledge the financial support from the Science and Engineering Research Board (SERB), Government of India, under the MATRICS grant. 
\section*{Code Availability}
The code used in this work is made available at \href{https://github.com/anup12352/AMMCR}{\textcolor{blue}{https://github.com/anup12352/AMMCR}}.
%\clearpage

\bibliography{reference}
\end{document}